\documentclass[12pt]{article}
\usepackage{amsmath,amssymb}
\usepackage{cite}

\makeatletter

\makeatletter
\def\eqnarray{%
  \stepcounter{equation}\def\@currentlabel{\p@equation\theequation}%
  \global\@eqnswtrue \m@th \global\@eqcnt\z@ \tabskip\@centering
  \let\\\@eqncr
  $$\everycr{}\halign to\displaywidth\bgroup
    \hskip\@centering$\displaystyle\tabskip\z@skip{##}$\@eqnsel
   &\global\@eqcnt\@ne \hfil$\;{##}\;$\hfil%          <--- [1]
   &\global\@eqcnt\tw@ $\displaystyle{##}$\hfil\tabskip\@centering
   &\global\@eqcnt\thr@@ \hb@xt@\z@\bgroup\hss##\egroup
    \tabskip\z@skip
    \cr}
\makeatother

\def\unit{{1\kern-.65ex {\rm l}}}
\def\1{{1\kern-.65ex {\rm l}}}
\def\ap{{\alpha'}}

\def\CI{{\cal I}}

\def\CM{{\cal M}}

\def\CO{{\cal O}}

\def\CR{{\cal R}}

\def\bbR{{\mathbb{R}}}

\newcount\hour \newcount\minute
\hour=\time \divide \hour by 60
\minute=\time
\def\now{%
\ifnum \hour<13
  \ifnum \hour=0 \advance \hour by 12 \number\hour:\else \number\hour:\fi%
     \ifnum \minute<10 0\fi%
     \number\minute%
\ A.M.%
\else \advance \hour by -12 \number\hour:%
  \ifnum \minute<10 0\fi%
  \number\minute%
  \ P.M.%
\fi%
}

\makeatother

\begin{document}

% format
\baselineskip=18pt  % a la harvmac
\numberwithin{equation}{section}  % make eq labels (sec.num)
\allowdisplaybreaks  % allow page breaks in displayed eqs

% print date, time and filename 
%\pagestyle{myheadings}
%\markright{{\tt \jobname.tex} -- \today{} \now}

%%%%%%%%%%%%%%%%%%%%%%%%%%%%%%%%%%%%%%%%%%%
%%%        TITLE BEGINS HERE
%%%%%%%%%%%%%%%%%%%%%%%%%%%%%%%%%%%%%%%%%%%

%% ========== title (note version) begins here ==========

%\vspace*{-1cm}
%\begin{center}
% {\Large\bf Notes on 4D Small Black Rings}
%\end{center}
%\vspace*{-.5cm}

%% ========== title (note version) ends here ==========

% ========== title (paper version, a la harvmac) begins here ==========

\thispagestyle{empty}

% Report number
\vspace*{-2cm} 
\begin{flushright}
{\tt arXiv:0710.4139}\\
NSF-KITP-07-187\\
ITFA-2007-47
\end{flushright}

% title, authors, affiliation
\vspace*{2.5cm} 
\begin{center}
 {\LARGE Are There Four-Dimensional Small Black Rings?}\\
% {\LARGE Are There Small Black Rings in Four Dimensions?}\\
 \vspace*{1.7cm}
 Norihiro Iizuka$^1$ and Masaki Shigemori$^2$\\
 \vspace*{1.0cm} 
 $^1$ 
{\it Kavli Institute for Theoretical Physics, \\
University of California, Santa Barbara, CA 93106-4030, USA \\[1ex]
 $^2$ 
Institute for Theoretical Physics, University of Amsterdam\\
Valckenierstraat 65, 1018 XE Amsterdam, The Netherlands\\}
 \vspace*{0.8cm} 
 \verb|iizuka_at_kitp.ucsb.edu|, 
 \verb|mshigemo_at_science.uva.nl|
\end{center}
\vspace*{1.5cm}

% abstract
\noindent

% short version
In $d>4$ dimensions, one can argue for the existence of small black
rings using a scaling argument.  We apply the same scaling argument to
the $d=4$ case and demonstrate that it fails to say anything about the
existence of $d=4$ small black rings, because stringy corrections get out of
control.
%For the scaling argument to work, there
%must exist a scaling region where i) the moduli dependence drops out of
%the solution, ii) the solution depends on the charge in a specific way,
%and iii) the curvature is smaller than the string scale.  In the $d=4$
%case the condition iii) cannot be satisfied.  
General relativity theorems say that there does not exist a black hole
with toroidal topology for $d=4$, but we interpret this as saying that,
for $d=4$ small black rings, stringy corrections are crucial which
invalidate the assumptions those theorems are based on.

%In $d>4$ dimensions, one can argue for the existence of supersymmetric
%small black rings using a scaling argument.  For this, it is crucial
%that there exist a scaling region in the SUGRA solution where i) the
%moduli dependence drops out of the solution, ii) the solution depends on
%the charge only in a specific way, and iii) the curvature is smaller
%than the string and Planck scales.  Once such a scaling region exists,
%then one can determine the charge-dependence of the solution at the
%horizon, assuming that higher derivative corrections create an event
%horizon.
%%
%We apply the same scaling argument to the $d=4$ case and find that there
%exists a region where i) and ii) are satisfied but iii) cannot be
%satisfied.  This demonstrates that higher derivative corrections must be
%more important for $d=4$, if there exist small black rings.
%%
%Topological censorship theorems say that there does not exist a black
%object with topology of torus in asymptotically flat four-dimensional
%spacetime, but we interpret this as saying that, for four-dimensional
%small black rings, stringy corrections are crucial which invalidate the
%assumptions those theorems are based on.

\newpage
\setcounter{page}{1} % don't number title page

% ========== title (paper version, a la harvmac) ends here ==========

%%%%%%%%%%%%%%%%%%%%%%%%%%%%%%%%%%%%%%%%%%%
%%%           TITLE ENDS HERE
%%%%%%%%%%%%%%%%%%%%%%%%%%%%%%%%%%%%%%%%%%%

%\tableofcontents
%\printindex

%%%%%%%%%%%%%%%%%%%%%%%%%%%%%%%%%%%%%%%%%%%
%%%        MAIN TEXT BEGINS HERE
%%%%%%%%%%%%%%%%%%%%%%%%%%%%%%%%%%%%%%%%%%%

\section{Introduction and Conclusion}

Black holes have always been fascinating objects which deepen our
understanding not only of general relativity but also of gauge theory
and string theory.  The discovery of the five-dimensional black ring
%with non-spherical topologies 
\cite{Emparan:2001wn} demonstrated that the uniqueness theorems of
general relativity can take very different forms in different
dimensions.\footnote{For recent progress in the studies of the phase
structure of higher dimensional black holes and ringsi (``blackfolds''),
see, {\it e.g.}, \cite{Emparan:2007wm}.}  On the other hand, recent
technologies \cite{Dabholkar:2004yr, Dabholkar:2004dq, Sen:2004dp,
Sen:2005pu, Sen:2005ch, Sen:2005kj, Dabholkar:2005by, Dabholkar:2005dt,
Kraus:2005zm, Kraus:2005vz} to incorporate higher-derivative corrections
predicted by string theory have enabled us to show the existence of
small black holes, which cannot exist in general relativity only with a
two-derivative action.

The existence of small black holes was predicted by Sen more than a
decade ago by using a scaling argument \cite{Sen:1995in}.  In
\cite{Dabholkar:2006za}, a similar scaling argument was applied to argue
for the existence of supersymmetric small black rings in $d \ge 5$ dimensions, which
goes as follows.  First, one constructs the solutions corresponding to 
``small black rings'', at the level of SUGRA\@.
These solutions do not have finite horizons and actually have naked
singularities.  However, in the region not too close to the singularity,
where curvature is much smaller than the string and Planck scales, the
SUGRA solution must be trustable. Next, in this region, near the
singularity, one finds the scaling region where {\bf i)} moduli
dependence drops out of all the fields, namely metric, $B$-field and
dilaton, and {\bf ii)} charge dependence enters in the solution only in
a certain combination, and {\bf iii)} the curvature is much smaller than
the string and Planck scales so the solution is trustable.
Property i) is expected due to the attractor mechanism
\cite{Ferrara:1995ih, Strominger:1996kf, Ferrara:1996dd, Ferrara:1996um,
Ferrara:1997tw, Sen:2005wa, Goldstein:2005hq}.
If one goes further near the singularity, then the curvature becomes
large and the condition iii) breaks down; stringy corrections are
important.  But because of the properties i) and ii), the higher
derivative corrections can change the SUGRA solution only in a specific
combination\footnote{
%One can also assign the charges in such a way that
%$g_s$ corrections are suppressed, due to properties i) and ii), so that
%only $\alpha'$ corrections are important.
By scaling the charges appropriately, one can make the dilaton as small
as one wants, using the properties i) and ii).  Therefore, $g_s$
corrections are suppressed and only $\alpha'$ corrections are important.
}, and as a result, assuming
that the higher derivative corrections create an event horizon, one can
predict the charge dependence of the horizon geometry. Then, using the
Bekenstein-Hawking-Wald relation, one can predict the entropy formula of
the small black ring, which indeed reproduces the correct charge
dependence of entropy predicted from the microscopic arguments, up to a
numerical constant factor.

%Because of the presence of higher derivative terms, we must really use
%Wald's entropy formula rather than the simple area-entropy relation, but
%putting this inessential point aside, this simple argument indeed gives
%the correct charge dependence of entropy predicted from the microscopic
%arguments.
 
In this article we apply the same argument to the small black ring in
asymptotically flat $d = 4$ spacetime to see whether it goes through in
this case or not.  Microscopically, this system is a fundamental
heterotic string with momentum charge, winding charge, and angular
momentum and, as far as the microscopic entropy counting is concerned,
there is no difference between $d = 4$ and $d > 4$ cases.
Macroscopically, on the other hand, although the argument is very
similar to the higher dimensional cases, we will show that there is one
crucial difference: in the region where all the conditions i)--iii) were
satisfied for $d>4$, there is one condition that fails.  Namely, in this
region, although i) and iii) still hold, ii) is not satisfied for $d=4$.
If we go further near the singularity, the condition ii) can be
satisfied, but now the condition iii) does not hold.  Therefore, for
$d>4$, there is no region where all i)--iii) are satisfied.

%for $d > 4$ there is a scaling region where i)--iii)
%all hold as we go near the ring singularity of the spinning string SUGRA
%solution, whereas for $d=4$ the region where i) and ii) hold does exist,
%but in that region iii) does \emph{not} hold.  

%$original$
%This suggests that, in $D>4$, even before we take into account the
%higher derivatives corrections, at the level of SUGRA, the properties of
%small black rings can be understood.  Stringy higher derivative
%corrections only modifies the geometry near the singularities to create
%horizon. But in 4D, the small black rings are not at all predicted at
%the level of SUGRA. This seems to show that as far as we consider
%supergravity theory and take into account the higher derivative
%correction perturbatively, then there is no way to construct small black
%ring in 4D.

This suggests the following: for $d>4$, the properties of small black
rings can essentially be understood at the level of SUGRA, even before we
take into account the higher derivative corrections.  The small black
ring spacetime is determined in the scaling region, and stringy higher
derivative corrections only modifies the geometry further near the
singularities to create horizons.
On the other hand, for $d=4$, there is no scaling region accessible by
SUGRA, and SUGRA fails to say anything about the existence of small
black rings. This seems to suggest that, as far as we start from SUGRA
and take into account higher derivative correction perturbatively, it is
impossible to construct small black rings in $d=4$.

In the next section, we give a more detailed analysis which leads to the
conclusion above.  In the following section, we discuss our result in
connection with the known no-go theorems of four-dimensional black
rings.

\section{Small black rings in four dimensions}

\subsection{$d$ dimensions}

Consider heterotic string in $\bbR_t\times \bbR^{d-1}\times S^1\times
T^{9-d}$ with $4\le d\le 9$ with $\bbR_t\times \bbR^{d-1}$ denoting
$d$-dimensional Minkowski space. Let the radius of $S^1$ be $R_d$, and
the volume of $T^{9-d}$ be $(2\pi)^{9-d}\ap^{(9-d)/2}$.
Our objects of interest are the BPS
elementary string excitations in this theory carrying $n$ units of
momentum and $-w$ units of winding charge along the circle $S^1$ and
angular momentum $J$ in a two dimensional plane of $\bbR^{d-1}$.  The
fundamental string looks like a ring in the $d$ dimensional spacetime
$\bbR_t\times \bbR^{d-1}$ and becomes a small black ring
\cite{Iizuka:2005uv, Dabholkar:2005qs} for $d=5$ and is expected to be so for $d\ge 6$.
These small black rings carry charges $n,w,J$ as well as charge $Q$
which corresponds to the winding number along the ring direction.

The metric for a small black ring for general $d$ is, at the two
derivative level \cite{Dabholkar:2006za, Emparan:2001ux},
\begin{eqnarray}
 ds^2_{str,d+1}&=&
  f_f^{-1}[-(dt-A_i dx^i)^2+(dx^d-A_i dx^i)^2+(f_p-1)(dt-dx^d)^2]
  + d{\bf x}_{d-1}^2
  \nonumber\label{newmetric_SBR_gen}\\
 e^{2\Phi_{d+1}}&=& g^2 \, f_f^{-1},\qquad
 B_{td}= -(f_f^{-1}-1),\qquad
  B_{ti}=-B_{di}= f_f^{-1}\, A_i,
\end{eqnarray}
where $i=1,2,\dots,d-1$, and
\begin{equation}
 \begin{split}
 f_f &= 1 + \frac{Q_f}{R^{d-3}} \left(\frac{x-y}{-2y}\right)^{(d-3)/ 2}
	    {}_2 F_1\left(\frac{d-3}{4},\frac{d-1}{4}; 1; 1-\frac{1}{y^2}\right),
  \\
 f_p &= 1 + \frac{Q_p}{R^{d-3}} \left(\frac{x-y}{-2y}\right)^{(d-3)/ 2}
	    {}_2 F_1\left(\frac{d-3}{4},\frac{d-1}{4}; 1; 1-\frac{1}{y^2}\right),
	    \label{newfffpA_xy}\\
 A_i dx^i &=-\frac{d-3}{2}\frac{q}{R^{d-5}}
            \frac{(y^2-1)(x-y)^{(d-5)/2}}{(-2y)^{(d-1)/2}}
	    \,{}_2 F_1\left(\frac{d-1}{4},\frac{d+1}{4}; 2; 1-\frac{1}{y^2}\right)
	    \, d\psi.
\end{split}
\end{equation}
The $(x,y)$ coordinate system is defined by (see Appendix \ref{app:xy}):
\begin{eqnarray}
 d{\bf x}_{d-1}^2
 &=& {R^2\over (x-y)^2}
  \left[
   {dy^2\over y^2-1}+(y^2-1)d\psi^2+{dx^2\over 
1-x^2}+(1-x^2)d\Omega_{d-4}^2
  \right].\label{defxy}
\end{eqnarray}
The range of the coordinates is $-1\le x\le 1$, $y\le -1$, $0\le\psi\le 2\pi$.

The various quantities are written in terms of microscopic quantities as
\begin{equation}
 q={16\pi G_d\over (d-3)\Omega_{d-2}\alpha'}Q,\quad
 R^2=\alpha'{J\over Q},\quad 
 Q_f={16\pi G_d R_d\over (d-3)\Omega_{d-2}\alpha'}w,  \quad
 Q_p={16\pi G_d\over (d-3)\Omega_{d-2}R_d}n\, .
  \label{newrel_qn_gen}
\end{equation}
Here $\Omega_D$ is the area of unit $D$-sphere
 and $G_d$ is the $d$-dimensional
Newton constant obtained by regarding the 
$S^1$ direction as compact,
\begin{eqnarray} \label{enewton}
 16\pi G_d&=&{16\pi G_{d+1}\over 2\pi R_d}
  ={(2\pi)^{d-3} g^2\ap^{(d-1)/2}\over R_d }.
\end{eqnarray}
$n,w,J,Q$ are integers.

The strategy of \cite{Dabholkar:2006za} was to find a ``scaling region''
in spacetime where
{
\addtolength{\abovedisplayskip}{1.5ex} 
\addtolength{\belowdisplayskip}{1.5ex} 
\begin{multline}
\begin{minipage}{0.9\textwidth}
\begin{itemize}
 \item[i)] all the moduli dependence drops out from the solution
	 (metric, dilaton, and $B$-field), and
 \item[ii)] charge dependence shows up only in specific combinations in
	 the solution, and
 \item[iii)] the curvature is small enough for the metric obtained from
	 the two-derivative action to be trustable.
\end{itemize}
\end{minipage}
\label{rescl_reg_conds}
\end{multline}
}
If one goes closer to the ring singularity, the condition iii) above
breaks down and the solution obtained by two-derivative action is no
longer trustable.  In this strongly coupled region, higher derivative
correction will modify the two-derivative solution, but it must do so in
a specific way determined by the specific charge combinations appearing
in the metric in the weakly curved region.  This allows one to
\emph{exactly} determine the charge dependence of the higher-derivative
corrected metric, up to some unknown functions.  That is enough to show
that the entropy derived from the Wald entropy formula agrees with the
microscopic entropy up to a factor.

The $d>4$ case was discussed in \cite{Dabholkar:2006za}, and it was
shown that the all conditions i)--iii) can indeed be met, if we go close
enough to the ring ({\it i.e.}, we take $|y|$ to be large enough), but
not too close ({\it i.e.}, we keep $|y|$ to be not too large) so that
the curvature is small.  We will see that for a $d=4$ small black ring,
if we go close enough to the ring so that i) and ii) are satisfied, the
last condition iii) is no longer satisfied.

\subsection{Four dimensions}

Let us focus on the $d=4$ case.  In this case there is no
$d\Omega_{d-4}^2$ term in \eqref{defxy}, and by defining $x=\cos\phi$,
\begin{eqnarray}
 d{\bf x}_{3}^2
 &=& {R^2\over (x-y)^2}
  \left[
   {dy^2\over y^2-1}+(y^2-1)d\psi^2+d\phi^2
  \right].
\end{eqnarray}
The harmonic functions are
%\begin{align}
% f_f
% &= 1 + \frac{Q_f}{R} \sqrt{\frac{x-y}{-2y}}
%	  \,  {}_2 F_1\left(\frac{1}{4},\frac{3}{4}; 1; 1-\frac{1}{y^2}\right)
% =1+{2Q_f\over \pi R}\sqrt{{x-y\over -2y\,\,}}\,\,
% {K\!\left({2\sqrt{1-y^{-2}}\over 1+\sqrt{1-y^{-2}}}\right) \over \sqrt{1+\sqrt{1-y^{-2}}}}\notag\\
% &=1+{Q_f\over \pi R}\left[
% \log|y|+3\log2+\CO\left({1\over|y|},{\log|y|\over|y|}\right)
% \right]
% \notag,\\
% f_p
% &=1+{Q_p\over \pi R}\left[
% \log|y|+3\log2+\CO\left({1\over|y|},{\log|y|\over|y|}\right)
% \right]
% \notag\\
% A&=-\frac{qR}{2}
%            \frac{(y^2-1)}{(x-y)^{1/2}(-2y)^{3/2}}
%	    \,{}_2 F_1\left(\frac{3}{4},\frac{5}{4}; 2; 1-\frac{1}{y^2}\right)
%	    \, d\psi.
%\end{align}
\begin{align}
 f_f
 &= 1 + \frac{Q_f}{R} \sqrt{\frac{x-y}{-2y}}
	  \,  {}_2 F_1\left(\frac{1}{4},\frac{3}{4}; 1; 1-\frac{1}{y^2}\right)
 =1+{2Q_f\over \pi R}\sqrt{{x-y\over -2y\,\,}}\,\,
 {K({2z\over z-1}) \over \sqrt{1-z}}\notag\\
 &=1+{Q_f\over \pi R}\left[
 \log|y|+3\log2+\CO\left({1\over|y|},{\log|y|\over|y|}\right)
 \right],
 \notag\\
 f_p
 &=1+{Q_p\over \pi R}\left[
 \log|y|+3\log2+\CO\left({1\over|y|},{\log|y|\over|y|}\right)
 \right],
 \notag\\
 A&=-\frac{qR}{2}
            \frac{(y^2-1)}{(x-y)^{1/2}(-2y)^{3/2}}
	    \,{}_2 F_1\left(\frac{3}{4},\frac{5}{4}; 2; 1-\frac{1}{y^2}\right)
	    \, d\psi\notag\\
 &=-{qR\over \pi}\sqrt{-2y\over x-y}
 {K({2z\over z-1})+(z-1)E({2z\over z-1}) \over \sqrt{1-z}},\notag\\
 &=-{4qR\over\pi}\left[\log|y|+3\log2-2+\CO\left({1\over|y|},{\log|y|\over|y|}\right)\right],
\end{align}
where $z\equiv \sqrt{1-1/y^2}$, and $K(m)$ and $E(m)$ are the complete
elliptic integrals of the first and second kinds, respectively:
\begin{align}
 K(m)&=\int_0^{\pi/2}\!\!\!{d\theta\over\sqrt{1-m\sin^2\theta}},\qquad
% =\int_0^1\!\!\!{dt\over\sqrt{(1-t^2)(1-mt^2)}}.
 E(m)=\int_0^{\pi/2}\!\!\!\sqrt{1-m\sin^2\theta\,}\,d\theta.
\end{align}
For large enough $|y|$ satisfying
\begin{align}
 \log|y|\gg {R\over Q_f},~{R\over Q_p},~1,\label{cond_y}
\end{align}
we can approximate these harmonic functions as
\begin{align} \label{easymp}
f_f \simeq   \frac{Q_f}{\pi R}  \log|y|,\qquad
f_p \simeq   \frac{Q_p}{\pi R}  \log|y|,\qquad
A_\psi \simeq
   - \frac{q R}{\pi}  \log|y|.
\end{align}

As was done for $d>4$ in \cite{Dabholkar:2006za}, let us consider the
following scaling of charges:
\begin{align}
 J\gg Q\gg 1,\qquad n\sim w,\qquad
 nw\sim JQ,\qquad 1-{JQ\over nw}\sim 1.\label{chgscl}
\end{align}
Then, using the relations \eqref{newrel_qn_gen}, \eqref{enewton}, we see
that the condition \eqref{cond_y} becomes\footnote{For comparison, the
corresponding relations for $d>4$ are
\begin{align}
 |y|^{d-4}\gg 
 \left({R\over\sqrt\ap}\right)^{d-4} {1\over g^2 Q},~
 \left({R\over\sqrt\ap}\right)^{d-4} {R_d^2\over g^2 Q\ap},~
 1.
\end{align}
}
\begin{align}
 \log|y|\gg {1\over g^2Q},~{R_d^2\over g^2 Q\ap},~1.\label{cond_y2}
\end{align}
This can certainly be met if we take $Q$ to be large.

The string frame, four-dimensional curvature goes for large $|y|$ as
\begin{align}
 \CR&\sim \left({|y|\log|y|\over R}\right)^2.
\end{align}
Therefore, for the metric \eqref{newmetric_SBR_gen}, which was derived
using the two derivative action, to be trustable, we need
\begin{align}
 |y|\log|y|\ll {R\over\sqrt\ap}.\label{cond_y3}
\end{align}
Because $|y|\gg 1$ from \eqref{cond_y2}, note that \eqref{cond_y3} implies
\begin{align}
 |y|\ll {R\over\sqrt\ap}.\label{cond_y3-1}
\end{align}

Using \eqref{easymp}, we see that for large $|y|$ satisfying \eqref{cond_y2},
the metric \eqref{newmetric_SBR_gen} takes the following form:
\begin{align}
 ds_{str,5}^2&={Q_p\over Q_f}(dx^4-dt)^2+{2\pi R\over Q_f \log|y|}dt(dx^4-dt)
 +2{qR^2\over Q_f}d\psi(dx^4-dt)\nonumber\\
 &\qquad+R^2{dy^2\over y^4}+R^2d\psi^2+{R^2\over y^2}d\phi^2,\qquad\\
 e^{2\Phi_5}&={\pi Rg^2\over Q_f\log |y|},\\
 B&=-{\pi R\over Q_f\log|y|}dt\wedge (dx^4-dt)
 -{R^2q\over Q_f}(dt-dx^4)\wedge d\psi+{\rm const.}
\end{align}
Or, in terms of microscopic numbers (Eq.\ \eqref{newrel_qn_gen}), the
metric is
\begin{equation}
\begin{split}
  ds_{str,5}^2&={n\ap\over w R_4^2}(dx^4-dt)^2+{4\pi\over g^2w}\sqrt{J\over Q}{1\over \log|y|}dt(dx^4-dt)
 +2{J\ap\over wR_4}d\psi(dx^4-dt)\label{metric_asym_num}\\
 &\qquad+R^2{dy^2\over y^4}+R^2d\psi^2+{R^2\over y^2}d\phi^2,\qquad\\
 e^{2\Phi_5}&={2\pi \over w}\sqrt{J\over Q}{1\over\log|y|},\\
 B&=-{2\pi \over wg^2}\sqrt{J\over Q}{1\over \log|y|}dt\wedge (dx^4-dt)
 -{J\ap\over wR_4}(dt-dx^4)\wedge d\psi+{\rm const.}
\end{split}
\end{equation}

Let us introduce the coordinates analogous to the ones introduced in
\cite{Dabholkar:2006za}:
\begin{align}
 \chi&\equiv \sqrt{J\over Q}\psi+\sqrt{JQ\over w}{1\over R_4}(x^4-t),\qquad
 \sigma\equiv {1\over R_4}\sqrt{{n\over w}-{JQ\over w^2}}(x^4-t),\\
 \tau&\equiv {2\pi R R_4\over\ap^{3/2}\sqrt{nw-JQ}g^2}t,\qquad
 \rho\equiv -{R\over \sqrt\ap\, y}.
\end{align}
%The way to obtain this is first to complete the square in $ds_{str,5}^2$
%with respect to $d\psi$ in \eqref{metric_asym_num}.  Then one sees that it's
%natural to introduce $\chi$ as defined above.  The rest are simply
%rescaling of coordinates to eliminate explicit dependence on charges.
The periodicity $(\psi,x^4)\cong (\psi+2\pi,x^4)\cong (\psi,x^4+2\pi
R_4)$ implies the following periodicity:
\begin{align}
 (\sigma,\chi)&\cong
 \left(\sigma,\chi+2\pi\sqrt{J\over Q}\right)
 \cong\left(\sigma+2\pi\sqrt{n\over w}\sqrt{1-{JQ\over nw}},\chi+2\pi{\sqrt{JQ}\over w}\right).
\end{align}
Also note that the third condition in \eqref{cond_y} and the condition
\eqref{cond_y3-1} become in terms of $\rho$ as follows:
\begin{align}
 1\ll\rho\ll{R\over\sqrt{\ap}}.\label{cond_rho}
\end{align}
In the new coordinate system, the metric \eqref{metric_asym_num} takes
the following simple form:
\begin{equation}
\begin{split}
  {ds_{str,5}^2\over\ap}&= d\chi^2+d\sigma^2+{2d\sigma d\tau\over\log({R\over \sqrt{\ap}\rho})}
 +d\rho^2+\rho^2d\phi^2,\\
 e^{2\Phi_5}&= {2\pi \over w}\sqrt{J\over Q}{1\over\log({R\over \sqrt{\ap}\rho})},\\
 {B\over\ap}&=-{1\over \log({R\over \sqrt{\ap}\rho})}d\tau\wedge d\sigma
 +{d\sigma\wedge d\chi\over \sqrt{nw/JQ-1}}.
\end{split}\label{nhmet_newcoord}
\end{equation}
In contrast with the $d>4$ case, we were unable to quite eliminate the
unwanted charge dependence from the metric in the region
\eqref{cond_rho} (recall that $R/\sqrt{\ap}=\sqrt{J/Q}$).
In order to eliminate the unwanted charge dependence from
\eqref{nhmet_newcoord} completely, we need to be able to approximate
\begin{align}
 \log({R\over \sqrt{\ap}\rho})
 =\log({R\over \sqrt{\ap}})+\log{1\over \rho}
 \stackrel{?}{\approx}\log{1\over \rho}.
\end{align}
This would mean
\begin{align}
 {R\over\sqrt\ap} \stackrel{?}{\ll} {1\over\rho}.\label{ivio25Jul07}
\end{align}
If this inequality were to hold, the charge dependence would enter only
in (the exponential of) dilation as an overall factor as one can see
from \eqref{nhmet_newcoord}.  Because (the exponential of) dilation
enters in the action as a overall factor at the string tree level, this
charge dependence would not influence the equation of motion.  However,
\eqref{cond_rho} means that $1\ll{R\over\sqrt\ap}$, which combined with
\eqref{ivio25Jul07} would mean
\begin{align}
 \rho\ll 1, \qquad \text{or}\qquad {R\over\sqrt\ap}\ll|y|.\label{largelargey}
\end{align}
This would contradict with \eqref{cond_rho} or \eqref{cond_y3-1}, the
condition for the curvature to be small.  Therefore, in this region, the
solution \eqref{nhmet_newcoord} obtained by two-derivative action is
actually not valid.  Namely, we could eliminate the unwanted charge
dependence, but for that we have come too close to the ring singularity
and the curvature, and thus the $\ap$ corrections, have gone out of
control.

Instead, one could stay in the region where the curvature is not very
strong and the solution \eqref{nhmet_newcoord} is valid.  In this
region, the conditions i) and iii) in \eqref{rescl_reg_conds} are
met, but the condition ii) is not completely satisfied.  But because the
solution \eqref{nhmet_newcoord} is trustable, we can start from it and
solve the $\ap$-corrected\footnote{Because dilaton is suppressed for
\eqref{chgscl}, we can use the heterotic string action at the string
tree level.  Although we do not know the string corrected action, it is
expected that the dilation gets small near a fundamental string even if
we included such corrections.} equations of motion toward smaller
$\rho$.  Then the solution must take the following form:
\begin{equation}
\begin{split}
  {ds_{str,5}^2\over\ap}&=g_{\alpha\beta}\left(\rho,{J\over Q}\right)d\zeta^\alpha d\zeta^\beta 
 +f_1\left(\rho,{J\over Q}\right)d\phi^2+d\rho^2,\\
 e^{2\Phi_5}&= {1 \over w}\sqrt{J\over Q}\,f_2\left(\rho,{J\over Q}\right),\\
 {B\over\ap}&=b_{\alpha\beta}\left(\rho,{J\over Q}\right)d\zeta^\alpha d\zeta^\beta
 +{d\sigma\wedge d\chi\over \sqrt{nw/JQ-1}}.
\end{split}
\end{equation}
where $g_{\alpha\beta},f_1,f_2$, and $b_{\alpha\beta}$ are some unknown
functions representing our ignorance of the $\ap$-corrected higher
derivative action. $\zeta^\alpha$ stands collectively for the
coordinates $\tau,\sigma$ and $\chi$.  Because those unknown functions
depend on $J/Q$, even if we assume that the $\ap$ corrections lead to a
finite horizon, we cannot determine the Wald entropy of the small black
ring as was possible for $d>4$ \cite{Dabholkar:2006za}.  Namely,
because the condition ii) is not satisfied, we do not have enough
control over the higher derivative corrections.

In summary, for $d=4$, there does not exist a region where all the
conditions i)--iii) in \eqref{rescl_reg_conds} are met.  If we go
very close to the ring (very large $|y|$ satisfying
\eqref{largelargey}), the charge dependence seems to disappear from the
metric \eqref{newmetric_SBR_gen}, but in that region the curvature has
become much larger than the string scale and the metric
\eqref{newmetric_SBR_gen} itself is not valid.
It is possible that there does exist a region where i)--iii) are all
satisfied, but it should be very near the ring where the curvature is of
the order of the string scale, and we need to know all $\ap$ corrections
in order to be able to study such a region.

\section{Discussion}

%% Original
%What is the relation of our result to the general relativity theorems
%\cite{Hawking:1971vc} on the topology of event horizons of
%four-dimensional black holes?  Based on topological censorship
%\cite{Friedman:1993ty}, following four-dimensional theorem is shown in
%\cite{Chrusciel:1994tr, Galloway:1999bp}.  Let $\CM$ be the spacetime,
%and assume that this spacetime can be conformally included into a
%spacetime-with-boundary, $\CM'=\CM\cup\CI$.  Then the topological
%censorship states that, every causal curve whose initial and final
%endpoints belong to $\CI$ is endpoint-homotopic to a curve in $\CI$.  As
%a result of this theorem, it was shown in \cite{Galloway:1999bp} that
%the sum of the genera of the event horizons is bounded by a number
%determined by the topology of $\CI$.  In particular, when the spacetime
%is asymptotically flat, this theorem shows that the topology of every
%black hole in four-dimension must have genus 0, and there cannot be a
%black ring whose horizon topology is a torus
%\cite{Chrusciel:1994tr,Galloway:1999bp}.

What is the relation of our result to the general relativity theorems
\cite{Hawking:1971vc} on the topology of event horizons of
four-dimensional black holes?  Based on topological censorship
\cite{Friedman:1993ty}, a powerful theorem about the topology of
four-dimensional black holes was proved in \cite{Chrusciel:1994tr,
Galloway:1999bp}.  Let $\CM$ be the spacetime, and assume that this
spacetime can be conformally included into a spacetime-with-boundary,
$\CM'=\CM\cup\CI$.  Then the topological censorship amounts to the
statement that every causal curve whose initial and final endpoints
belong to $\CI$ is endpoint-homotopic to a curve in $\CI$.  Based on
this topological censorship, it was shown in \cite{Galloway:1999bp} that
the sum of the genera of event horizons is bounded by a certain number
determined by the property of $\CI$.  In particular, when the spacetime
is asymptotically flat, this theorem says that the horizon of every
black hole must have genus 0, and there cannot be a black ring with
toroidal topology \cite{Chrusciel:1994tr,Galloway:1999bp}.

One might think that this is consistent with the result we found in this
note, and that there does not exist a small black ring in four
dimensions.
%even in the presence of higher derivative terms, 
%as long as higher derivative corrections are small enough.  
Note however that topological censorship is under certain
assumptions\footnote{See, for example, page 8 of
\cite{Galloway:1999bp}.}, which use notions of classical geometry based
on point particles.  In string theory, the fundamental object has a
finite size and the validity of such classical notions cannot be taken
for granted particularly when the curvature of spacetime is of the order
of the string scale. Therefore for the objects such as small black holes
or rings, it is logically possible that this theorem cannot be applied.

One related fact is that, in string theory, there is redefinition
ambiguity in the metric field:
\begin{align}
 g_{\mu\nu}&\to g_{\mu\nu}+a\,\ap R_{\mu\nu}+b\,\ap R g_{\mu\nu}+\dots,
\label{metric_redef}
\end{align}
which does not change the action up to two-derivative terms.  However,
under \eqref{metric_redef}, geodesic length of a curve, which is an
invariant quantity in classical gravity, changes.  In particular, such
an effect is not negligible if the curvature of spacetime is of the
order of string scale.  Therefore, in such situations, arguments based
on topological censorship must be reconsidered.

Furthermore, in string theory we know examples where the notion of
metric itself loses its sense.  Consider NS5-branes in flat
10-dimensional space.  Away from the NS5-branes, one can trust the SUGRA
metric \cite{Horowitz:1991cd}, but as one approaches the NS5-brane,
dilaton becomes large and one cannot trust the solution any more; in
this strongly coupled region, the metric is not a good variable to
describe the physics.  However, in this case we know what to do: we need
to go to the $S$-dual picture where one has D5-branes instead of
NS5-branes and the low energy physics is described by the dual metric
variable.  The original metric and dual metric are good variables in
different regions in spacetime, and theorems about the original metric
(in the region of its validity) has little to say about the dual metric
(in the region of its validity).  Therefore, it is conceivable that,
whatever general relativity theorems say about the metric away from the
singularity of the small black ring, in the strongly coupled region very
near the ring a dual metric becomes more appropriate, whose horizon
topology those theorems have nothing to say about.

To summarize, we showed that the scaling argument that could be used for
$d>4$ small black rings to derive their entropy formula is not valid for
the $d=4$ small black ring.  This is because for $d=4$ there does not
exist a scaling region where all the conditions i)--iii) in
\eqref{rescl_reg_conds} are met, which existed for $d>4$.  General
relativity theorems say that there does not exist a black hole with
toroidal topology in asymptotically flat four-dimensional spacetime, but
we interpret this as saying that stringy corrections are crucial for the
four-dimensional small black ring which invalidates the assumptions
underlying those theorems.

\section*{Acknowledgments}

We thank Jan de Boer, Kostas Skenderis, Marika Taylor and Erik Verlinde
for discussions.  We would also like to thank Atish Dabholkar, Ashik
Iqubal and Ashoke Sen for the collaboration in \cite{Dabholkar:2006za}
and for useful comments on the manuscript. Our special thanks go to
Roberto Emparan for fruitful discussions and for bringing our attention
to the relevant references.  The research of N.I. was supported in part
by the National Science Foundation under Grant No.\ PHY05-51164.  The
research of M.S. was supported by an NWO Spinoza grant.

\appendix

\section{$x,y$ coordinates}
\label{app:xy}

Denote the Cartesian coordinates for $\bbR^{d-1}$ by ${\bf
x}_{d-1}=(x^1,x^2,x^3,\dots,x^{d-1})$.  The relation of these
coordinates to the $(x,y,\psi,\Omega_{d-4})$ coordinates are as follows:
\begin{align}
 x^1=u\cos\psi,\quad x^2=u\sin\psi,\qquad
 (x^3,x^4,\dots,x^{d-1})=(v\xi^1,\dots,v\xi^{d-3}),
\label{xd_uvpsixi}
\end{align}
where $u,v\in[0,\infty)$, $\psi\in[0,2\pi)$ and
$(\xi^1,\dots,\xi^{d-3})$ are coordinates of $S^{d-4}$ satisfying
$(\xi^1)^2+\cdots+(\xi^{d-3})^2=1$.  Then the relation to the $(x,y)$
coordinate is \cite{Emparan:2006mm}
\begin{align}
 u&={\sqrt{y^2-1}\over x-y}R,\qquad
 v={\sqrt{1-x^2}\over x-y}R,\\
 x&=-{u^2+v^2-R^2\over \Sigma},\quad
 y=-{u^2+v^2+R^2\over \Sigma},\quad
 \Sigma=\sqrt{(u^2+v^2+R^2)^2+4R^2v^2}.
\end{align}
The flat metric for $\bbR^{d-1}$ is
\begin{align}
 d{\bf x}_{d-1}^2&=du^2+u^2d\psi^2+dv^2+v^2d\Omega_{d-4}^2\notag\\
 &= {R^2\over (x-y)^2}
  \left[
   {dy^2\over y^2-1}+(y^2-1)d\psi^2+{dx^2\over 1-x^2}+(1-x^2)d\Omega_{d-4}^2
  \right].
\end{align}

Note that, for $d=4$, $S^{d-4}$ is made of two points: $\xi^1=\pm 1$,
and  \eqref{xd_uvpsixi} becomes
\begin{align}
 x^1=u\cos{\psi},\qquad x^2=u\sin{\psi},\qquad x^3=\pm v.
\end{align}

In terms of the $(x,y)$ coordinates, the computation of harmonic functions
goes e.g.\ as
\begin{align}
 f_f&=
  1+{Q_f\over L}\int_0^L{dv\over |{\bf x}-{\bf F}(v)|^{d-3}}\notag\\
 &=1+{Q_f\over 2\pi}\left({x-y\over -2R^2y}\right)^{(d-3)/ 2}
 \int_0^{2\pi}{d\theta\over(1-\sqrt{1-y^{-2}}\cos\theta)^{d-3\over 2}}\notag\\
 &= 1 + Q_f \left(\frac{x-y}{-2R^2y}\right)^{(d-3)/ 2}
	    {}_2 F_1\left(\frac{d-3}{4},\frac{d-1}{4}; 1; 1-\frac{1}{y^2}\right).
\end{align}
See the appendix of \cite{Dabholkar:2006za} for more details.

%%%%%%%%%%%%%%%%%%%%%%%%%%%%%%%%%%%%%%%%%%%
%%%        MAIN TEXT ENDS HERE
%%%%%%%%%%%%%%%%%%%%%%%%%%%%%%%%%%%%%%%%%%%

\end{document}